\documentclass[twocolumn,showpacs,preprintnumbers,amsmath,amssymb,prl]{revtex4}

\usepackage{graphicx}
\usepackage{dcolumn}

\usepackage{amsmath,amssymb}

\usepackage[mathscr]{euscript}

\usepackage{calc}

\usepackage{nicefrac}

\usepackage{color}

\usepackage{mathptmx, courier, pifont}
\usepackage[scaled=0.92]{helvet}
\usepackage[T1]{fontenc}
\usepackage{textcomp} 

\usepackage{bm}

\newcommand{\singlefig}[6]{%
\begin{figure}[tpb]\vspace{#3}%
\includegraphics*[scale=#5]{#2}%
\caption{\label{fig:#1} #6}%
\vspace{#4}%
\end{figure}}

\newcommand{\doublefig}[6]{%
\begin{figure*}[tpb] \vspace{#3}%
\includegraphics*[scale=#5]{#2}%
\caption{\label{fig:#1} #6}
\vspace{#4}
\end{figure*}}

\newtheorem{conjecture}{Conjecture}

\newcommand{\tsub}[1]{_{\mbox{\scriptsize#1}}}

\newcommand{\bra}[1]{\langle#1|}
\newcommand{\ket}[1]{|#1\rangle}
\newcommand{\ev}[1]{\langle#1\rangle}
\newcommand{\mel}[3]{\bra{#1}#2\ket{#3}}

\newcommand{\spin}{\mathscr{S}}   

\newcommand{\shellfill}{f}

\newcommand{\kdegen}{\Omega_k}

\newcommand{\Piop}[2]{\Pi_{#1#2}}

\newcommand{\phantomdagger}{^{\vphantom{\dagger}}}

\newcommand{\sufour}{{\rm SU(4)}}
\newcommand{\sutwo}{{\rm SU(2)}}
\newcommand{\sofive}{{\rm SO(5)}}
\newcommand{\sosix}{{\rm SO(6)}}
\newcommand{\soseven}{{\rm SO(7)}}
\newcommand{\soeight}{{\rm SO(8)}}

\newcommand{\casimir}[1]{C_{\scriptscriptstyle#1}}

\newcommand{\eq}[1]{Eq.~(\ref{#1})}

\newcommand{\eqnoeq}[1]{(\ref{#1})}
\newcommand{\fig}[1]{Fig.~\ref{fig:#1}}

\newcommand{\mwgalign}[1]
{\hspace*{-0.4em}&#1&\hspace*{-0.4em}}

\newcommand{\halfthin}{\kern 0.0834em}
\newcommand{\neghalfthin}{\kern -0.0834em}

\newcommand{\quarterthin}{\kern 0.0417em}
\newcommand{\negquarterthin}{\kern - 0.0417em}

\newcommand{\Adag}[2]{A^\dagger_{#1#2}}

\newcommand{\hws}{\ket{\rm HW}}

\newlength{\figup}
\newlength{\tabup}
\newlength{\tabtitlesep}
\newlength{\afterTableLineOne}
\newlength{\afterTableLineTwo}
\newlength{\beforeTableLineThree}
\newlength{\tablerowMore}
\newlength{\tableLineShift}
\newcommand{\runningheads}[2]{\markboth{\hfill #1\hfill}{\hfill #2\hfill}}

\begin{document}

\title{The Ground State of Monolayer Graphene in a Strong Magnetic Field}

\author{Lian-Ao Wu$^{(1)}$}
\email{lianaowu@gmail.com}
\author{Mike Guidry$^{(2)}$}
\email{guidry@utk.edu}

\affiliation{
$^{(1)}$IKERBASQUE, Basque Foundation for Science, 48011 Bilbao, Spain,
and Department of Theoretical Physics and History of Science,
Basque Country University (EHU/UPV), Post Office Box 644, 48080 Bilbao, Spain
\\
$^{(2)}$Department of Physics and Astronomy, University of
Tennessee, Knoxville, Tennessee 37996, USA
}

\date{\today}

\begin{abstract} 
Graphene SU(4) quantum Hall symmetry is extended to SO(8), permitting analytical 
solutions for graphene in a magnetic field that break SU(4) spontaneously. We 
recover standard graphene SU(4) physics as one limit, but find new phases and 
new properties that may be relevant for understanding the ground state.  The 
graphene SO(8) symmetry is found to be isomorphic to one that occurs extensively 
in nuclear structure physics, and very similar to one that describes 
high-temperature superconductors, suggesting deep mathematical connections among 
these physically-different fermionic systems. 
\end{abstract}

\pacs{
73.22.Pr, 
73.43.-f  
}

\maketitle

\runningheads{}{{\em The Ground State of Monolayer Graphene in a 
Strong Magnetic Field}---L.-A. Wu and M. W. Guidry}

Graphene in a strong magnetic field has approximate SU(4) symmetry 
\cite{nomu2006,alic2006,yang2006,goer2006,khar2012,wufe2014,barl2012,youn12}, 
which permits examining explicit symmetry breaking by small 
terms in the  Hamiltonian. However, the ground state  is  strongly insulating, 
with rapid divergence of longitudinal resistance at a critical magnetic 
field  $B\tsub c$ \cite{chec2008}.   The dependence of $B\tsub c$  on sample 
impurities suggests that the resistance is an intrinsic property of an {\em 
emergent state} differing qualitatively from perturbed SU(4) solutions 
(spontaneous symmetry breaking) \cite{jung2009}. SU(4) symmetry can suggest 
the form of possible emergent states but cannot describe them quantitatively. 
Numerical simulations find various possible ground states  having similar 
energies but differing structure.  Thus the nature of the insulating ground 
state remains elusive.

Here we show that SU(4) symmetry can be extended to an SO(8) symmetry that 
recovers graphene SU(4) physics, but that implies  new low-energy modes that 
transcend SU(4) symmetry and for which solutions may be obtained {\em 
analytically.}  As a first application  we  revisit the nature of the ground 
state for undoped monolayer graphene in a magnetic field.

Good reviews of graphene physics are available 
\cite{cast2009,goer2011,barl2012};  we recall here only features relevant for 
the present discussion. Graphene is bipartite with sublattices  A and B; the 
quantity specifying whether an electron is on the A or B sublattice is termed 
the {\em sublattice pseudospin}.  The dispersion computed in  
tight-binding approximation  \cite{cast2009,goer2011} indicates  two 
inequivalent sets of points  in the Brillouin zone, labeled $K$ 
and $K'$.  The two-fold K degree of freedom is  termed {\em valley isospin}. 
Near these K-points the dispersion is linear, leading to {\em Dirac cones.} 
For undoped graphene the Fermi surface lies at the apex of the cones, where the 
level density vanishes and the effective electronic mass tends to zero. Hence, 
near the K points low-energy electrons for undoped graphene in zero magnetic 
field obey a massless Dirac equation and  behave mathematically as {\em massless 
chiral fermions,} with chirality related 
to projection of the sublattice 
pseudospin.

In a magnetic field the massless Dirac equation may be solved with an 
appropriate vector potential and the resulting Landau levels (LL) are labeled by 
integers.  The $n=0$ level is unusual in that  it is half filled in the ground 
state of undoped graphene, leading to the anomalous counting observed in the 
graphene quantum Hall effect \cite{novo2005,zhan2005}. For low-energy 
excitations in each valley ($K$ or $K'$),  inter-valley tunneling may be ignored 
and the electrons in the valley reside entirely on either the A or B sublattice, 
implying that for the $n=0$ LL valley isospin  and sublattice pseudospin are 
equivalent labels.  We shall be concerned primarily with this $n=0$ Landau 
level, which has, in addition to the Landau orbital degeneracy, a 4-fold 
degeneracy corresponding to spin and valley isospin.

The largest energy scales are the LL separation and Coulomb energy.  For neutral 
graphene the LL separation is approximately three times larger than the Coulomb 
energy, which is in turn much larger than other interactions.  Hence, we shall 
ignore inter-LL excitations and consider only a single $n=0$ LL. Justification 
and caveats for this  approximation are discussed in Ref.\ 
\cite{khar2012}. We adopt a Hamiltonian  \cite{khar2012,wufe2014}%
\begin{equation}
H =
H_ 0 - H\tsub z +
\tfrac12 \sum_{i\ne j}\left[
 g_z\tau_z^i \tau_z^j + g_{\perp} (\tau_x^i \tau_x^j + \tau_y^i \tau_y^j)
 \right]
 \delta(\bm r_i \hspace{-1.5pt}-\hspace{-1.5pt} {\bm r}_j) 
\label{so5_1.1a}
\end{equation}
where the Pauli matrices $\tau_\alpha$ operate on  valley isospin, the Pauli 
matrices $\sigma_\alpha$ operate on  electronic spin, $g_z$ and $g_\perp$ are 
coupling constants, $\mu\tsub B$ is the Bohr magneton, and the spin $z$ 
direction is assumed aligned with the magnetic field.  The three 
terms in \eq{so5_1.1a} represent the valley-independent Coulomb interaction, the 
Zeeman energy, and the short-range valley-dependent interactions, respectively.

The four internal states representing possible combinations of the projection of 
the spin $\sigma$ and the projection of the valley isospin $\tau$ are displayed 
in \fig{grapheneBasis_withTable}.%
\singlefig
{grapheneBasis_withTable}       
{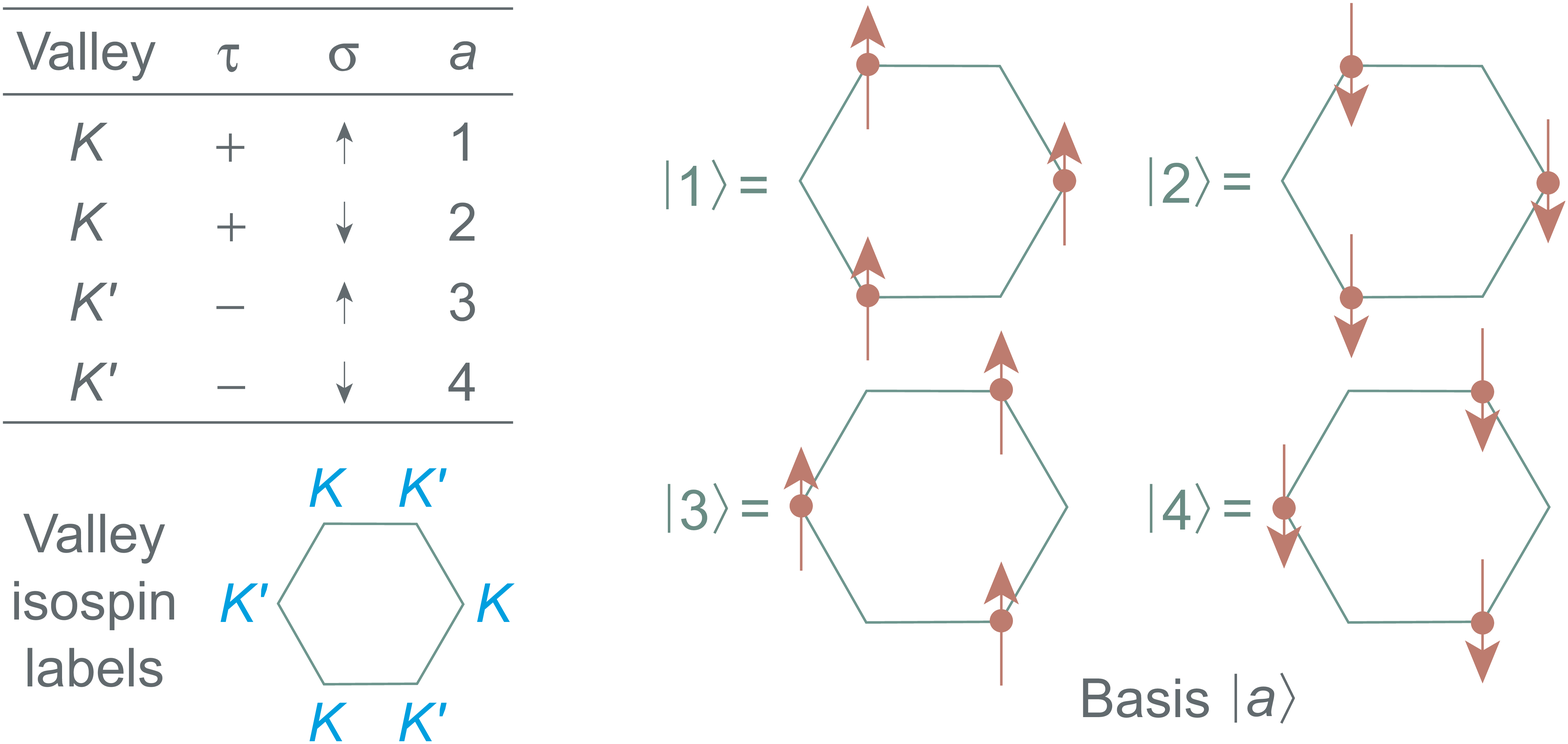}   
{\figup}         
{0pt}         
{0.16}         
{Isospin--spin quantum numbers and basis vectors.}
Symmetries of the Hamiltonian \eqnoeq{so5_1.1a} may be investigated by 
introducing the 
15 operators
\begin{subequations}
\begin{eqnarray}
 \spin_\alpha \mwgalign=
 \sum_{m_k}
\sum_{ \tau \sigma \sigma'} \mel{\sigma'}{\sigma_\alpha}{\sigma}
c^\dagger_{\tau\sigma'm_k} c_{\tau\sigma m_k}
\label{algebra1.4a}
\\
T_\alpha \mwgalign=
\sum_{m_k}\sum_{\sigma\tau\tau'} \mel{\tau'}{\tau_\alpha}{\tau}
c^\dagger_{\tau'\sigma m_k} c_{\tau\sigma m_k}
\label{algebra1.4b}
\\
N_\alpha \mwgalign=
\tfrac12 \sum_{m_k}\sum_{\sigma \sigma' \tau }
\mel{\tau}{\tau_z}{\tau}
\mel{\sigma'}{\sigma_\alpha}{\sigma} 
c^\dagger_{\tau\sigma' m_k}
c_{\tau\sigma m_k}
\label{algebra1.4c}
\\
\Piop\alpha\beta \mwgalign= 
\tfrac12 \sum_{m_k}\sum_{ \sigma \sigma' \tau \tau'}
\mel{\tau'}{\tau_\beta}{\tau}
\mel{\sigma'}{\sigma_\alpha}{\sigma}
c^\dagger_{\tau' \sigma' m_k}
c_{\tau\sigma m_k} 
\label{algebra1.4d}
\end{eqnarray}
\label{algebra1.4}%
\end{subequations}
where $c^\dagger(c)$  create (annihilate) fermions, $\alpha=(x, y, z)$, 
$\beta=(x, y)$, $\tau$ and $\sigma$ are defined in 
\fig{grapheneBasis_withTable}, and $m_k$ labels orbitally-degenerate LL states. 
Physically, $\spin_\alpha$ is  total spin,  $T_\alpha$ is  total valley 
isospin, $ N_\alpha$ is a N\'eel vector measuring the difference in spins on 
the A and B sublattices, and the  $\Piop\alpha\beta$ couple  spin and valley 
isospin. Under commutation the operators \eqnoeq{algebra1.4} close  an SU(4) 
algebra that commutes with the Coulomb interaction \cite{khar2012,wufe2014}. 
If terms 2 and 3 in \eq{so5_1.1a} are small compared with the first, the 
Hamiltonian has  approximate SU(4) invariance.  Explicit breaking of SU(4) 
depends on the values of $g_z$ and $g_\perp$.  Four symmetry-breaking patterns 
have been discussed  \cite{khar2012,wufe2014}.

For a 2-$N$ dimensional fermionic space  the most general bilinear products 
$c^\dagger_i c_j$ of creation--annihilation operators and their hermitian 
conjugates generate an SU($2N$) Lie algebra under commutation. Adding the  most 
general pair  operators $c_i^\dagger c^\dagger_j$ and $c_ic_j$  extends  
SU($2N$)  to  SO($4N$) \cite{unitaryAlgebra,wyb74}. The extended symmetry 
permits defining a (collective) subspace of the full Hilbert space spanned by 
products of pair creation operators acting on the pair vacuum.  An effective 
Hamiltonian constructed from a polynomial in the Casimir invariants of all 
groups in the subgroup chains of SO($4N$)  will then represent the most general 
Hamiltonian for the collective subspace, and will be diagonal in the subspace 
basis for each dynamical symmetry  chain.  Thus, the manybody problem can be 
{\em solved exactly} in the symmetry limits defined by each  subgroup chain 
\cite{FDSM}, and analytically in coherent-state approximation 
\cite{arec1972,gilm1972,pere1972,gilm1974,zhan90} otherwise. This approach has 
been applied extensively to strongly-correlated fermions in  various fields; for 
representative examples see \cite{clwu86,clwu87,FDSM,guid01,lawu03}.

For graphene we assume a single $n=0$ LL with  creation operators 
$c^\dagger_{\tau \sigma m_k}$ and hermitian conjugates $c_{\tau \sigma m_k}$.  
Degeneracy of the LL is denoted by $2\Omega$.  Accounting for  4-fold 
spin--valley degeneracy,
$
2\Omega = 4(2\kdegen) = 4BS/(h/e),
$
where $2\Omega_k$ is the LL  orbital degeneracy, $B$ is magnetic field, and $S$ 
sample size. The {\em fractional occupation} of the LL is
$
\shellfill \equiv n/2\Omega,
$
where $n$ is  electron number, and the  {\em 
filling factor} is
$
\nu = 4(\shellfill - \tfrac12).
$

Now we add to the 15 SU(4) generators of \eq{algebra1.4} the charge operator 
$S_0 = \tfrac12 (n-\Omega)$, the 6 pairing operators $S^\dagger$ and 
$D_\mu^\dagger (\mu = 0, \pm 1, \pm 2)$, and their 6 hermitian conjugates, with
\begin{equation}
\begin{array}{c}
S^\dagger 
= \tfrac{1}{\sqrt2} ( A_{14}^\dagger - A_{23}^\dagger )
\quad\ \
D_0^\dagger = \tfrac{1}{\sqrt2} ( A_{14}^\dagger + A_{23}^\dagger 
)
\\[8pt]
D_2^\dagger =  A^\dagger_{12}
\quad\ \
D_{-2}^\dagger =  A_{34}^\dagger
\quad\
D_1^\dagger = A_{13}^\dagger
\quad\ \
D_{-1}^\dagger =  A_{24}^\dagger 
\end{array}
\label{coupled1.4}
\end{equation}
where $A^\dagger_{ab}$ creates a pair of electrons, one in the 
$a=(\tau_1,\sigma_1)$ level and one in the $b=(\tau_2,\sigma_2)$ level, with the 
total $m_k$ of each pair coupled to zero term by term:
$
A^\dagger_{ab}  = \sum_{m_k} c^\dagger_{a m_k} c^\dagger_{b 
-m_k} .
$
We also introduce for later use the linear combinations
\begin{equation}
Q_\pm^\dagger\equiv 
 \tfrac12 (S^\dagger \pm D_0^\dagger ) .
 \label{Qdef}
\end{equation}
The 28 operators 
$\{ \spin_\alpha,\,  T_\alpha,\,  N_\alpha,\, \Piop \alpha x, \,\Piop \alpha y, 
\, S_0, \, S, \,S^\dagger, \,D\phantomdagger_{\mu},\, D^\dagger_\mu \}$
generate an SO(8) algebra with a graphene SU(4) 
subalgebra.  The full structure for SO(8) subgroup chains is given in 
\fig{chains_graphene_narrow}.%
 \singlefig
{chains_graphene_narrow}  
{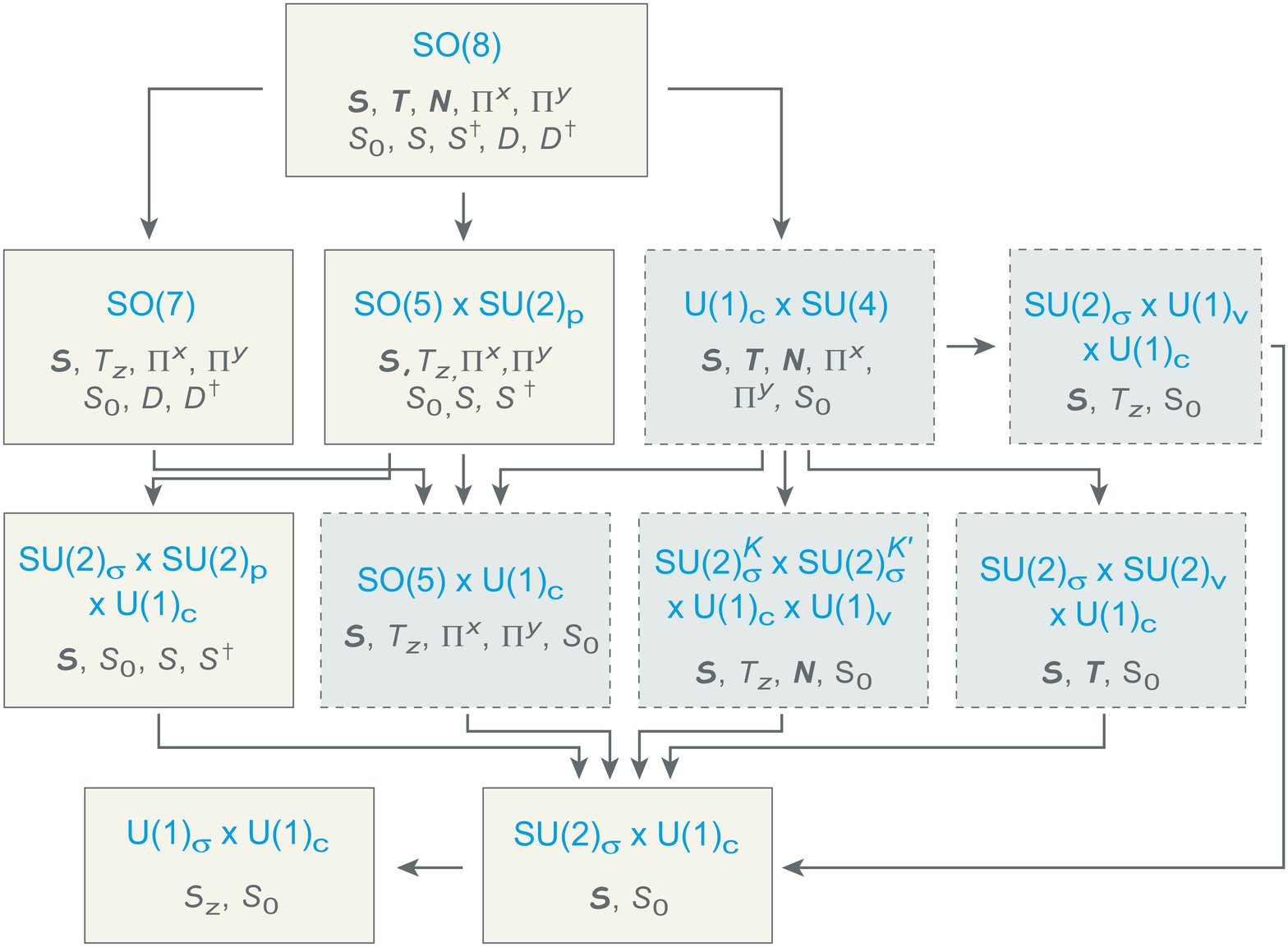}
{\figup}
{0pt}
{0.235}
{SO(8) subgroup chains with group generators. Dashed boundaries and darker 
shading indicate subgroups defining  the SU(4) quantum Hall 
model.  We see that SO(8) subsumes  graphene SU(4) but has a richer 
structure with additional subgroup chains.
}

Pair configurations created by generators of Eqs.\ \eqnoeq{coupled1.4} 
and \eqnoeq{Qdef} operating on the pair vacuum are given in 
\fig{S_D_pairs_brillouin}.%
\singlefig
{S_D_pairs_brillouin}       
{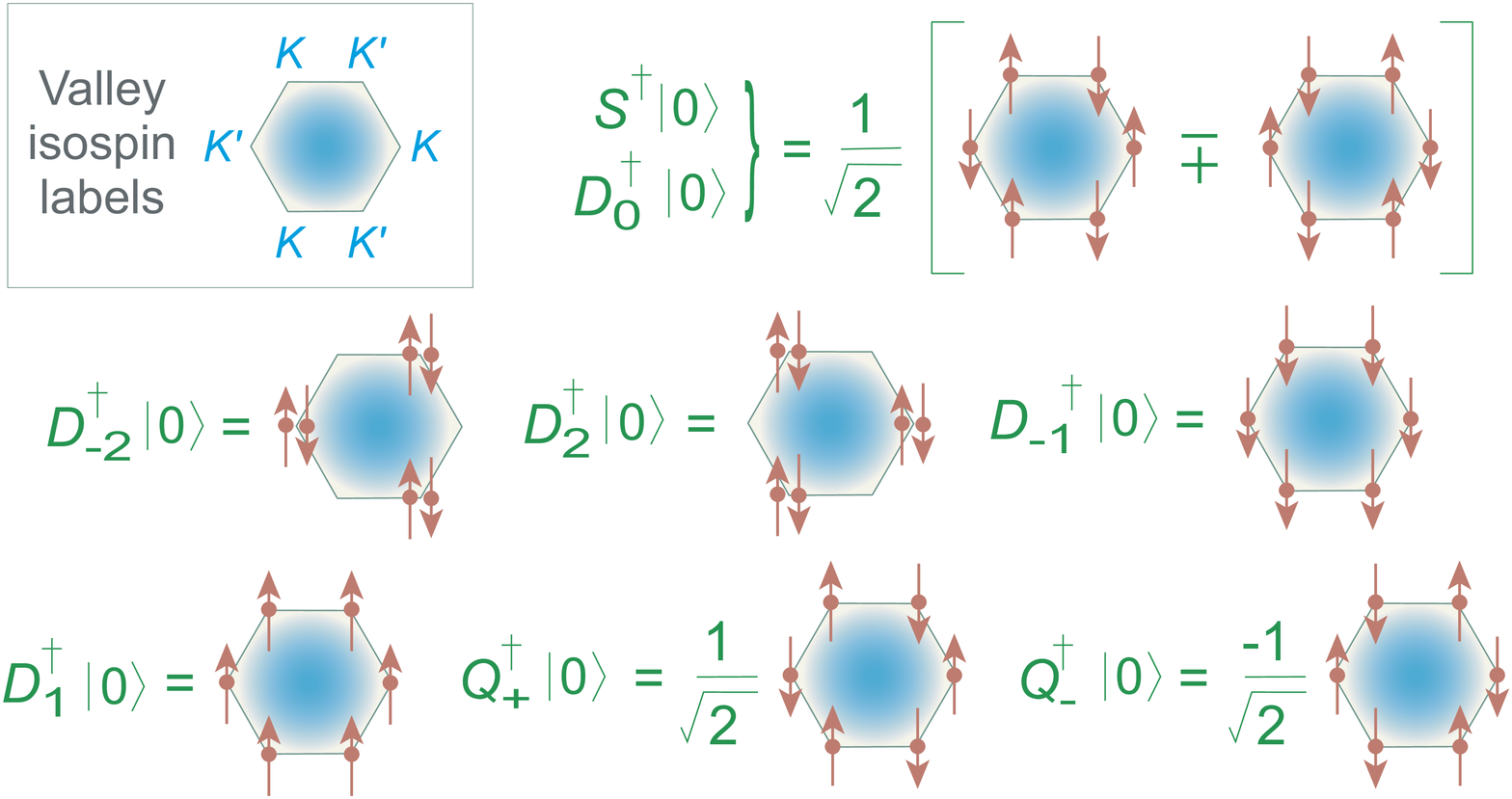}    
{\figup}         
{0pt}         
{0.19}         
{Configurations  created by the 
operators of Eqs.\ \eqnoeq{coupled1.4} and \eqnoeq{Qdef} operating on the pair 
vacuum $\ket0$.    Location of the dots ($K$ or $K'$ site) indicates the valley 
isospin; arrows indicate the spin polarization.}
Kharitonov \cite{khar2012} has classified collective modes for the $n=0$ LL 
using pairs  similar physically to ours: $D^\dagger_{\pm 2} \ket0$ creates 
spin-singlet charge density waves (CDW), $D^\dagger_{\pm 1} \ket0$ creates 
ferromagnetic (FM) states, and $Q_\pm^\dagger\ket0$ creates antiferromagnetic 
(AF) states. These states may be classified according to  $\ev{\spin_z}$, which 
measures net spin and characterizes {\em FM order,} $\ev{T_z}$, which measures 
the sublattice charge difference and characterizes {\em CDW order,} and 
$\ev{N_z}$, which measures the sublattice spin difference and characterizes {\em 
AF order.} Thus pairs in Eqs.\ \eqnoeq{coupled1.4}--\eqnoeq{Qdef} define  
modes already discussed \cite{khar2012,wufe2014}, but now 
SO(8) symmetry permits {\em analytical solutions} for corresponding collective 
modes. 

Broken-symmetry states based on graphene SU(4)  have been expressed in terms of 
the pair wavefunction \cite{khar2012}
\begin{eqnarray}
\ket\psi \mwgalign=
\prod_{m_k} 
\sum_{\tau\sigma\tau'\sigma'} \Phi^*_{\tau\sigma\tau'\sigma'}
c_{\tau\sigma m_k}^\dagger c_{\tau' \sigma' m_k}^\dagger
 \ket0 ,
\label{upairs1.8}
\end{eqnarray}
where  the product is over the LL orbital degeneracy label $m_k$ and the sum is 
over spin and valley labels. Now let us consider  SO(8) pairs. All states of an 
irreducible representation may be constructed by successive application of 
raising and lowering operators to a highest-weight (HW) state  (Cartan--Dynkin 
method) \cite{wyb74}. Let $u$ denote the number of broken pairs.  For $u=0$ 
states at half filling the pair number is
$
N=\tfrac12 \Omega = 2k+1
$
and the U(4) representation is  $(\tfrac\Omega2, \tfrac\Omega2, 0, 0)$.  We 
choose the HW state as the pair state that results from placing one electron in 
the $a=1$ and one  in the $a=2$ basis states (see 
\fig{grapheneBasis_withTable}),
\begin{eqnarray}
\ket{\rm HW} \mwgalign=
\frac{1}{N!} \left( \Adag12 \right)^{N} \ket0
= \frac{1}{N!}  \Big( \sum_{m_k}c^\dagger_{1 m_k}
c^\dagger_{2,-m_k}
\Big)^{N} \ket0 ,
\label{hwstate}
\end{eqnarray}
where the sum runs over the $N$ states in the LL labeled by the $m_k = (-k, 
-k+1, \ldots, k-1, k)$ orbital quantum number.  Writing the sum over $m_k$ in 
\eq{hwstate} out explicitly and invoking antisymmetry  eliminates most terms and 
leaves
\begin{eqnarray}
\ket{\rm HW} 
\mwgalign=
\frac{1}{N!}
\Big(\sum_{m_k} c^\dagger_{1m_k} c^\dagger_{2 -m_k}\Big)^{N}
\ket0 
= \hspace{-3pt}
 \prod_{m_k = -k}^{m_k=+k} c^\dagger_{1m_k} c^\dagger_{2m_k} \ket 0 .
\label{u0pairs1.4}
\end{eqnarray}
Thus the SO(8) $u=0$ HW state  is {\em equivalent to a product of pairs,} one  
for each $m_k$ in the LL.

Other states can be constructed by applying successively to  $\ket{\rm HW}$ 
ladder operators that are functions  of the generators $G = \{\spin_\alpha, 
T_\alpha, N_\alpha, \Pi_{\alpha\beta} \}$ of \eq{algebra1.4}.  For for an 
arbitrary state $\ket\psi$ in the weight space
$
\ket\psi = F(G) \ket{\rm HW}  ,
$
where $F(G)$ is specified by the Cartan--Dynkin procedure. For example, applying 
the isospin lowering operator $ T_- = F(G) \equiv \tfrac12 (T_x - iT_y)$ gives
\begin{eqnarray}
\ket{\psi} \mwgalign=
T_- \hws =
\prod_{m_k} 
\left(
c_{3 m_k}^\dagger
c_{2 m_k}^\dagger
+
c_{4 m_k}^\dagger
c_{1 m_k}^\dagger
\right)
\ket0 
.
\label{u0pairs1.7}
\end{eqnarray}
Likewise,  all other states of the $u=0$ representation can be constructed by 
using successive applications of raising and lowering operators fashioned from 
the generators of Eqs.\ \eqnoeq{algebra1.4}, and they will take the product of 
sums form \eqnoeq{u0pairs1.7}, just as for \eq{upairs1.8}.
\newcommand{\tuner}{q}
\doublefig
{coherentEnergySurfaces_composite}  
{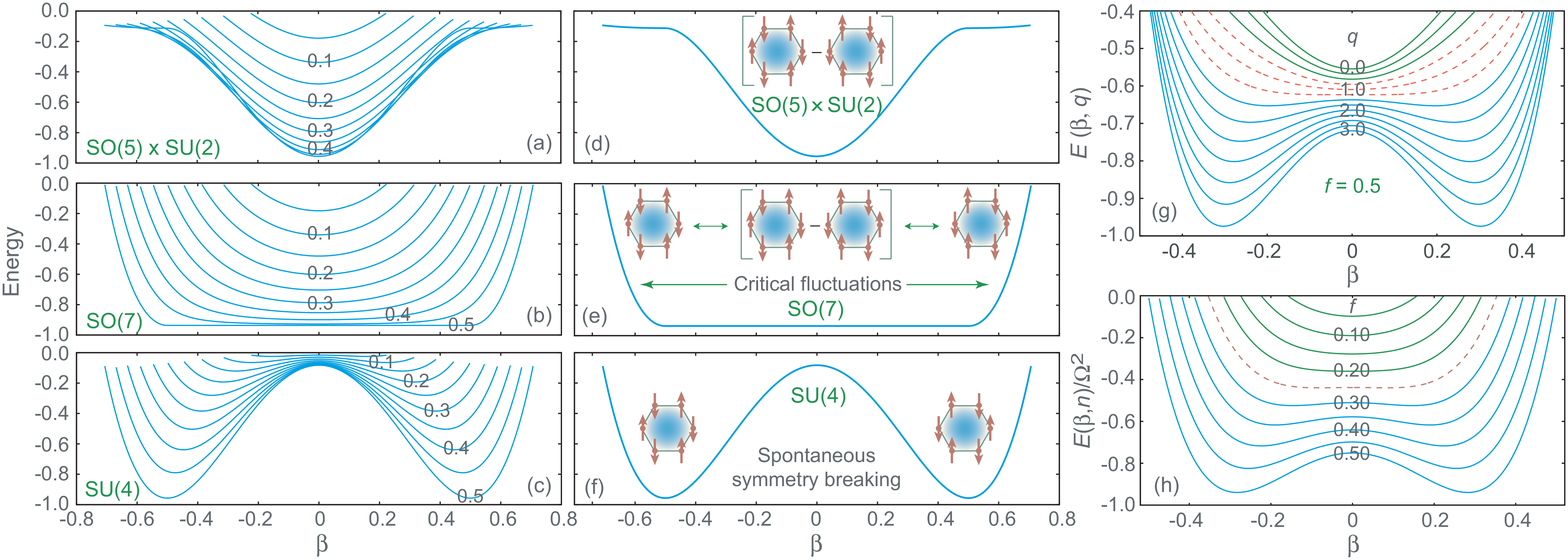}
{\figup}
{0pt}
{0.36}
{(a)--(c)~Coherent state energy surfaces as a function of AF order $\beta$ for 
three  SO(8) dynamical symmetry limits.  Curves labeled by  fractional 
occupation $\shellfill=n/2\Omega$ (particles or holes, since the SO(8) theory 
is particle--hole symmetric). (d)--(f)~Ground-state  energy   for three 
symmetry limits. Diagrams indicate  the wavefunctions suggested by Eqs.\ 
\eqnoeq{csWFs}. (g)~Energy as a function of coupling strength ratio $\tuner 
\equiv G_0/b_2$ for $\shellfill  = 0.5$. Solid green curves ($\tuner \sim 
0-0.5$) indicate  $\sofive\times\sutwo$ symmetry, solid blue curves for $\tuner 
\ge 1.5$ indicate SU(4) symmetry.  Dashed red curves for $\tuner \sim 1$ 
correspond to the critical SO(7) symmetry mediating the quantum phase transition 
from $\sofive\times\sutwo$ to SU(4). (h)~Energy surfaces for different 
occupation fractions $f$ at fixed  $G_0$ and $b_2$ ($q=2.5$). Solid green curves 
for $f \sim 0-0.2$ indicate $\sofive\times\sutwo$ symmetry, Solid blue curves 
for $f \sim 0.3-0.5$ indicate SU(4) symmetry.  Curves near $f \sim 0.25$ (dashed 
red) correspond to  SO(7) symmetry mediating the 
$\sofive\times\sutwo\leftrightarrow\sufour$ quantum phase transition.}
Hence the   $\soeight\supset\sufour$ symmetry introduced here recovers  existing 
understanding  \cite{khar2012,wufe2014} of  states  expected from spontaneous 
breaking of SU(4) by short-range correlations. 

Figure \ref{fig:chains_graphene_narrow} contains 7 subgroup chains. Each defines 
a dynamical symmetry realized for specific choices of the SO(8) Hamiltonian 
parameters, and  yields an {\em exact manybody solution} using standard 
techniques.  We shall deal with these exact solutions in future papers.  Here, 
we interpret the states implied by \fig{chains_graphene_narrow} using coherent 
state (CS) approximations \cite{zhan90}.  The 
full coherent state solution will be presented in a later paper, but we 
illustrate here for subgroup chains containing SO(5)  in 
\fig{chains_graphene_narrow}, corresponding to solutions that are linear 
combinations of the symmetry-limit solutions for the $\soeight \supset \sufour 
\supset \sofive \supset \sutwo$, the $\soeight \supset \sofive \times 
\sutwo\tsub p \supset \sofive \supset \sutwo$, and the $\soeight \supset 
\soseven \supset \sofive \supset \sutwo$ dynamical symmetries. The required 
group theory is already known
\cite{FDSM,gino80,chen86,zhan90,weimin1989,zhang88,wmzha87}, so we can (with 
suitable change of notation and basis) simply transcribe many  equations  and 
reinterpret them in terms of graphene physics. Details follow in a later paper 
but we show here  results central to this paper.

{\em Energy surfaces:} The energy $E\tsub g(n,\beta)$ depends only on  $n$ and a 
{\em single order parameter} $\beta$. For the symmetry groups g
\begin{equation}
E\tsub g(n,\beta) = N\tsub g \left[ A\tsub g\beta^4 + B\tsub g(n)\beta^2 + 
C\tsub g(n) + D\tsub g(n,\beta) \right] ,
\label{generalEnergyFormula}
\end{equation}
where the group-dependent coefficients are tabulated in Ref.\ \cite{zhang88}. 
This energy surface results from minimizing 
\begin{equation}
\ev H \simeq 
G_0\ev{\casimir{\sutwo\tsub p}}   + b_2\ev{\casimir{\sufour}}  ,
 \label{modelHamiltonian}
\end{equation}
where $\ev \ $ is taken in the CS,  $\casimir{g}$ denotes the quadratic Casimir 
invariant of  g, and $b_2$ and $G_0$ are coupling strengths.

{\em Order Parameter:}
The order parameter $\beta$
distinguishes the phases associated with the subgroup 
chains in \fig{chains_graphene_narrow} that contain SO(5). $\beta$ measures 
AF, since 
it is related to the  AF order parameter $\ev{N_z}$  by
$
\ev{N_z} =
2 \Omega |b_2| (\shellfill - \beta^2)^{1/2} \ \beta .
$

{\em Fluctuations:} Coherent states violate translational, rotational, and gauge 
invariance.  However, for realistic fields and sample sizes these violations 
are negligible, yielding a Ginzburg--Landau type theory with  
microscopic pedigree.

{\em Wavefunctions:}
Closed forms are given for the SO(8) CS wavefunctions in Ref.\ \cite{zhang88}. 
Evaluating these expressions in the  $\sofive\times\sutwo$ and SU(4) limits, 
respectively, gives 
\begin{equation} 
\ket{{\rm SO}_5\times{\rm SU}_2} \simeq ( S^\dagger )^N \ket0
\qquad
\ket{{\rm SU}_4} \simeq
 ( Q_\pm^\dagger )^N  \ket0 .
 \label{csWFs}
\end{equation}
Thus the SU(4) state is a superposition of $Q_\pm$  pairs, each with vanishing 
$\ev{\spin_z}$ and $\ev{T_z}$ but finite AF order  $\ev{N_z}$.  Conversely, the 
$\sofive\times\sutwo$ state  is a superposition of $S$  pairs, each with 
vanishing $\ev{\spin_z}$, $\ev{T_z}$, and $\ev{N_z}$. The critical SO(7) state 
is realized in the transition from $\sofive\times\sutwo$ to SU(4) and represents 
a complex mixture of these wavefunctions.

Energies for the $\sofive\times\sutwo$, SO(7), and SU(4)  limits are shown for 
several values of $\shellfill = n/2\Omega$ in 
\fig{coherentEnergySurfaces_composite}(a)--(c). The solutions are distinguished 
by the AF order parameter $\beta$ at the minimum, which is zero for 
$\sofive\times\sutwo$, non-zero for SU(4), and indeterminate in the SO(7) 
critical dynamical symmetry that interpolates  between $\sofive\times\sutwo$ and 
SU(4) through  fluctuations in  AF order. For undoped graphene the ground state  
corresponds to the $\shellfill = 0.5$ curves. These are shown in 
\fig{coherentEnergySurfaces_composite}(d)--(f) for the three symmetry limits, 
along with a physical interpretation of the states in terms of the wavefunctions 
\eqnoeq{csWFs}. Thus the SO(8) dynamical symmetry limits illustrated in 
\fig{coherentEnergySurfaces_composite}(d)--(f) represent a rich set of 
collective states that can be distinguished by the expectation value and 
fluctuations associated with the order parameter $\beta$.

Quantum phase transitions between symmetry limits may be studied by varying 
coupling. We rewrite \eq{modelHamiltonian} in terms of a parameter $\tuner 
\equiv b_2/G_0$ favoring  $\sofive\times\sutwo$ when $\tuner << 1$, SU(4)  when 
$\tuner >> 1$,  and  SO(7) when $\tuner \sim 1$ [$ \ev{\casimir{\sutwo\tsub p}}  
 + \ev{\casimir{\sufour}} \sim \ev{\casimir{\soseven}} $, implying SO(7) 
symmetry if $G_0 \sim b_2$]. Variation of ground state energy  with  $\tuner$  
is shown in \fig{coherentEnergySurfaces_composite}(g). Alternatively, at fixed  
$q$ phase transitions  may be initiated by changing particle occupancy. Figure 
\ref{fig:coherentEnergySurfaces_composite}(h) displays a transition from 
$\sofive\times\sutwo$  with $\beta=0$, through a critical SO(7) symmetry with 
energy  highly degenerate in $\beta$, to SU(4)  with $\beta \ne 0$,  as $f$ is 
changed at constant $q$.

Thus SO(8) describes analytically a host of broken-SU(4)  candidates for the 
states in graphene being unraveled in modern experiments 
\cite{youn12,feld2012,benj2013,youn14}.  These solutions provide a spectrum of 
excited states as well as ground states.  We shall not discuss that here, except 
to note that all ground state solutions have a gap to electronic and collective 
excitations. The general  theory to be discussed in forthcoming papers can 
accommodate FM, CDW, and AF states, but  for dynamical symmetries containing  
SO(5) all solutions may be classified by a {\em single parameter} $\beta$ that 
measures AF order:  SU(4) states have finite $\beta$ and AF order, but no CDW or 
FM order, $\sofive\times\sutwo$ states have $\beta=0$ and no AF, CDW, or FM 
order, and SO(7) states define a critical dynamical symmetry that interpolates 
between SU(4) and $\sofive\times\sutwo$ with no AF order but  large AF 
fluctuations, and with no CDW or FM order. We have neglected Zeeman coupling 
here but it is expected to be small  for the $n=0$ LL \cite{youn12}, primarily  
leading to AF canting \cite{khar2012}.

Transport properties are not manifest in the algebraic theory but the CS 
approximation is equivalent to symmetry-constrained 
Hartree--Fock--Boboliubov (HFB) theory \cite{zhan90,weimin1989}, suggesting 
that  SO(8) theory can be mapped onto Hartree--Fock (HF) transport calculations. 
HF calculations for armchair nanoribbons found that AF and CDW states 
similar to ours have no edge currents \cite{jung2009}. 
We speculate that our AF states also are insulating and thus strong 
candidates for the high-field ground state, but confirmation requires more 
work. 

Solutions depend on $G_0$ and $b_2$  in  \eq{modelHamiltonian}, which define 
effective interactions in the truncated space [highly renormalized relative to  
parameters in \eq{so5_1.1a}]. They may be fixed by systematic comparison with 
data, enabling a robust prediction for the nature of the  ground and other 
low-energy states. We expect modest
impurity levels to modify the effective interaction parameters but leave 
dynamical symmetries intact.

The present ideas are similar to ones found in nuclear physics \cite{FDSM} and 
high-$T\tsub c$ superconductors (SC) \cite{guid01,lawu03}, with all three cases 
exhibiting  $\soeight\supset\sufour \ [\sim \sosix]$  symmetry and a {\em 
critical dynamical symmetry} generalizing a quantum critical 
point to a {\em quantum critical phase} linking other phases through quantum 
fluctuations. In graphene and nuclear SO(8) the critical symmetry is SO(7), 
which interpolates between $\sofive\times\sutwo$ and $\sufour$ \cite{wmzha87}; 
in SC it is SO(5), which interpolates between SU(2) SC and SO(4) AF Mott 
insulators \cite{guid01,lawu03}. These similarities may have implications for 
cross-disciplinary understanding of quantum phase transitions. 

In summary, we have introduced an SO(8) model of monolayer graphene in a 
magnetic field that recovers SU(4) quantum Hall physics but implies new 
collective modes transcending explicitly-broken SU(4) that are leading 
candidates for the high-field ground state. Graphene SO(8) is isomorphic to a 
symmetry describing many complex nuclei 
and very similar to one describing high-$T\tsub c$ superconductors, suggesting a 
deep mathematical connection among these phenomena.

We thank Yang Sun and his Shanghai students for useful discussions.  L. W. 
acknowledges grant support from the Basque Government Grant No.\ IT472-10 and 
the Spanish MICINN Grant No.\ FIS2012-36673-C03-03.  This work was partially 
supported by LightCone Interactive LLC.

\bibliographystyle{unsrt}


\end{document}